\documentclass{llncs}
\usepackage{graphicx} 
\usepackage{cite}

\usepackage[many]{tcolorbox}
\definecolor{main}{HTML}{CFCFCF}  
\definecolor{sub}{HTML}{CFCFCF}   

\tcbset{
    sharp corners,           
    colback = white,         
    before skip = 0.2cm,     
    after skip = 0.5cm       
}

\newtcolorbox{boxC}[2][]{aibox,title=#2,#1}
\tcbset{
  aibox/.style={
    top=10pt,
    boxrule=0.5pt,
    rounded corners,
    coltitle=black,
    colframe=gray,
    colbacktitle=sub,
    colback = sub,
    enhanced,
    center,
    attach boxed title to top left={yshift=-0.1in,xshift=0.15in},
    boxed title style={boxrule=0.5pt,colframe=gray,},
  }
}

\title{Temporal Network Analysis of Microservice Architectural Degradation}
\author{Alexander Bakhtin\orcidID{0000-0003-3513-7253}}
\institute{University of Oulu, Oulu, Finland \\  \email{{alexander.bakhtin}@oulu.fi}}

\begin{document}

\maketitle

\begin{abstract}
Microservice architecture can be modeled as a network of microservices making calls to each other, commonly known as the service dependency graph. Network Science can provide methods to study such networks. In particular, temporal network analysis is a branch of Network Science that analyzes networks evolving with time.  In microservice systems, temporal networks can arise if we examine the architecture of the system across releases or monitor a deployed system using tracing.

In this research summary paper, I discuss the challenges in obtaining temporal networks from microservice systems and analyzing them with the temporal network methods. In particular,
the most complete temporal network that we could obtain contains 7 time instances and 42 microservices, which limits the potential analysis that could be applied.
\end{abstract}


\section{Introduction}
Microservice architecture is a paradigm promoting the decomposition of a coupled software system into small, independent components, called microservices \cite{lewis14_microservices}.
Many patterns and anti-patterns were proposed in the literature to improve the design and architecture of microservice systems \cite{cerny2023catalog}.
Over time, the architecture of a microservice system can decay. Such a phenomenon is known as microservice architectural degradation.

We can model the composition of microservices, i.e., their architecture and their interactions, as networks. Such networks can be studied as static or evolving over time, i.e., so-called \emph{temporal} networks, leveraging techniques and methods from Network Science \cite{al2022using, bakhtin2025network}.
In the context of microservice systems, temporal networks arise when considering either a network of a microservice system taken for each release or commit in the project history \cite{bakhtin2025ccp}, or a real-time network of monitoring traces collected from a running system \cite{al2022using}.
Our aim is to leverage such representation to detect architectural degradation and provide feedback to the architects.

\section{Related Work}
In this section, I discuss the related work leveraging network methods for the analysis of microservice architecture and its degradation,
although such methods have not been widely utilized for this problem. Moreover, most of the current approaches do not leverage the temporal dimension of the data.

For instance, Cerny et al. \cite{cerny2023catalog} conducted a tertiary study on architectural anti-patterns for microservices and provided a comprehensive catalog. 
The \textit{topology} sub-category represents the only case where anti-patterns may be detected leveraging the microservice architectural network.
Current approaches for detecting such anti-patterns
either consider individual call chains, or
utilize the aggregated, static network, where the order and frequency of the calls are ignored \cite{al2022using}. Analyzing the entire temporal network of traces could provide a more detailed and fine-grained insight into these anti-patterns \cite{guo2020graph}. For instance, Saarimäki et al. \cite{saarimaki2022towards} highlight the importance of incorporating time as a critical yet underexplored dimension in software analysis.

Furthermore, Cerny et al. \cite{cerny2025towards} have argued for adopting an evolutional study of microservice systems to assess change impact, which is currently done in monolithic projects. They argue that current approaches and tools focus on studying each microservice individually, which can miss instances when a change in one service can have an impact on other services in the system \cite{robredo2024analyzing}.
Notably, they identified works that deal with co-evolution analysis and evolution graphs  \cite{lelovic2024change} and showcased an evolution analysis of the \emph{train-ticket} benchmark system \cite{abdelfattah2024assessing}
However, the authors did not perform any kind of network analysis on the reconstructed microservice architectures, while Abufouda and Abukwaik \cite{abufouda2017using} argued for properly adopting temporal network methods in Empirical Software Engineering.
In our recent paper, accepted in the current edition of ECSA \cite{bakhtin2025ccp}, we demonstrated the temporal analysis of \emph{train-ticket} leveraging temporal centrality, while the paper from the previous edition \cite{bakhtin2024temporal} leveraged temporal community detection for developer collaboration networks.

\section{Proposed Methodology}
We can leverage diverse types of temporal network methods for analysis of microservice systems with different goals.
In this section, I consider how we can analyze two types of microservice networks presented in the introduction: microservice architectural evolution and microservice real-time tracing.

\subsection{Microservice Architectural Evolution}

To be able to perform temporal architectural analysis, we require a suitable microservice system and reconstructed architecture \cite{amoroso2024dataset, bakhtin2023tools, schneider2024comparison_preprint} available for many versions of the system \cite{bakhtin2024challenges}.

Temporal centrality analysis \cite{taylor, bakhtin2025ccp} can provide a metric indicating the importance of each node temporally. If certain nodes become too important and central, this can be an indication that they are becoming bottlenecks and critical points in the system \cite{cerny2023catalog, al2022using}. We have recently performed such kind of analysis on the \emph{train-ticket} microservice benchmark \cite{bakhtin2025ccp} and found that the centrality metric does not correlate with size, complexity, or quality metrics, thus being potentially a novel metric to consider \cite{bakhtin2025network}, or an orthogonal perspective.

Moreover, temporal community detection \cite{gauvin2014detecting} can uncover groups of nodes that are tightly connected to each other, potentially highlighting services that exchange too many calls and are too highly coupled. For instance, Gauvin et al. \cite{gauvin2014detecting} proposed a community detection method that can demonstrate in which time instances such a group was present, even if the time instances are not consecutive, thus shedding light on the re-emergence of highly coupled groups. Indeed, we have previously applied their method on a network of developer collaboration in an open-source (OSS) microservice benchmark \cite{bakhtin2024temporal}. We discovered that each release was prepared by a single tightly coupled group of people, with only one core developer being present during the entire development and a couple of others for most of the development.

Finally, we also aim to investigate if disease spreading simulations \cite{newman2002spread}, such as Susceptible-Infected-Susceptible \cite{lee2013epidemic}, can be applied to the temporal network \cite{holme2016temporal} of microservice architecture versions given some quality attributes, such as the presence of bugs, code smells, or violations \cite{esposito2024validate}, to see if their presence spreads like a disease once they start affecting one service \cite{robredo2024analyzing, esposito2023uncovering}.

\subsection{Microservice Real-time Tracing}

Microservice tracing data is an important aspect of monitoring microservice systems \cite{bakhtin2025lo2}. Tracing data allows the practitioner/researcher to see invocation chains of services and analyze how different services and their latency or stability affect other services and the health of the overall microservice system \cite{al2022using}. Tracing data can be analyzed leveraging temporal network methods developed for online processing, i.e., consuming and analyzing incoming data in real-time.

In particular, temporal centrality has the potential to be calculated and updated in real-time. Unlike the algorithm by Taylor et al. \cite{taylor}, which requires the availability of the entire temporal network at once,  Beres et al. \cite{beres2018temporal} provided an algorithm that computes the centrality scores of nodes in a temporal network given as a stream of connections. The scores are updated whenever a new connection is registered. Therefore, we aim to investigate if the real-time centrality score can function as a novel metric for microservice monitoring by correlating microservice centrality with latency or other metrics \cite{bakhtin2025network}. The centrality score could highlight problematic services that receive too many requests and thus need to be scaled-up or potentially refactored into several services.

Similar to temporal community detection, it is possible to detect different system states to identify anomalies. Masuda and Holm \cite{masuda2019detecting} provided a method to cluster the different time intervals of a temporal network to identify intervals when a network appears to be in a certain state and when this state changes.
A possible research direction would be to observe whether the detected states of a microservice system can be identified with the anomalies in the system given the ground truth of the induced anomalies, or correlate the state with some other metrics observed in the system. Leveraging this method could provide a light-weight, purely trace-based method for microservice anomaly detection.

Finally, we can focus on detecting groups of nodes that stay tightly connected for a certain interval of time, known as cliques. Thus, mining cliques in the temporal network is a process similar to community detection. Lin et al. \cite{lin2021mining} provide an algorithm to detect such structures that we can leverage for a network of microservice traces. In a network of microservice monitoring, such a clique would indicate a group of services that exchange a lot of communication consistently, thus indicating the potential presence of architectural anti-patterns such as Inappropriate Service Intimacy, Wrong Cuts, or Microservice-greedy system \cite{cerny2023catalog}.

\section{Encountered Challenges}
In the current stage of my PhD work and research, I have encountered challenges when pursuing both of the directions outlined in this work. The challenges are related to the availability of Industrial and OSS microservice systems and their data. They are detailed below.

\subsection{Microservice Architectural Evolution}

As part of our group's research work, I investigated the availability of OSS microservice systems suitable for analysis \cite{amoroso2024dataset} and tools for microservice architecture reconstruction \cite{bakhtin2023tools}, as well as comparison and evaluation of such tools leveraging static analysis \cite{schneider2024comparisonrr, schneider2024comparison_preprint}. 
However, an attempt to combine the most promising tool prototype, \emph{Code2DFD} \cite{schneider2023automatic}, and all applicable projects from the identified dataset, i.e., Java Spring Projects, resulted in only 24 projects having a reliably reconstructed architectural network \cite{bakhtin2025network}.
This result, together with the results of the tool evaluation \cite{schneider2024comparison_preprint}, poses the following challenge:

\noindent\textbf{Challenge 1} \textit{Lack of microservices systems with a provided architectural diagram or which can be reconstructed in detail with a reliable tool.}

After identifying the 24 projects for which reliable architecture was obtained for at least the latest version \cite{bakhtin2025network}, I also attempted to reconstruct the previous versions of the projects in order to obtain the temporal architectural network. However, since most of the projects are small example benchmarks, they do not show variations in architecture \cite{bakhtin2024challenges}. The only suitable project, i.e., the project with enough components, connections, and clear architectural evolution, turned out to be \emph{train-ticket} \cite{zhou2018benchmarking}, which is the de facto standard subject of research on microservices anyway \cite{al2022using, bakhtin2022microservice, zhang2022tracecrl, avritzer2025architecture, zhao2025does}. Thus, it was the only project we could analyze in our most recent paper \cite{bakhtin2025ccp}, inevitably affecting the validity and generalizability of the study. Furthermore, even the temporal network of \emph{train-ticket} only contains 7 time instances and 42 microservices. This poses the following challenge:

\noindent\textbf{Challenge 2} \textit{Lack of microservice systems for which variations in architecture over time are known.}

Access to a big industrial system would, of course, provide a system that has evolved significantly. However, it is not guaranteed that its architecture has been precisely documented in all releases or that it could be reconstructed across the history.

\subsection{Microservice Real-time Tracing}
Collecting tracing data through dynamic analysis of the system is a laborious and time-consuming task, which requires some load scenarios to be generated and executed on the running system deployed with sufficient resources \cite{al2022using}. As such, data of the necessary scale and with realistic scenarios and distributions is only in the hands of industrial companies with massive microservice systems that do not publish it openly.

Several "tech giants", such as eBay \cite{guo2020graph}, Alibaba \cite{luo2021characterizing}, and IBM \cite{pourmajidi2019dogfooding}, published research on analysis and monitoring of their proprietary microservice systems; however, due to industry practices, they do not share the analyzed data.
This situation by itself poses a certain challenge:

\noindent\textbf{Challenge 3} \textit{Lack of access to Industrial-scale datasets of microservice traces.}

Conversely, most research leveraging OSS benchmark projects \cite{al2022using, bakhtin2022microservice, zhang2022tracecrl, avritzer2025architecture, zhao2025does} seems to rely on either \emph{train-ticket} \cite{zhou2018benchmarking} or \emph{DeathStarBench} \cite{gan2019open} projects, which appear to be the two biggest available projects in terms of the number of microservices. Others include \emph{SockShop}\footnote{\url{https://github.com/microservices-demo/microservices-demo}}, which is significantly smaller in scope, and \emph{OnlineBoutique}\footnote{\url{https://github.com/GoogleCloudPlatform/microservices-demo}}, which requires deployment onto the Google Cloud platform.
Such concentrated attention on a handful of benchmarks created by researchers poses a threat to validity and generalizability of works leveraging them, thus posing a challenge:

\noindent\textbf{Challenge 4} \textit{Lack of diverse OSS microservice benchmarks that can be efficiently leveraged for dynamic trace analysis.}

As part of the \emph{Multimodal Fusion-based Anomaly Detection in Microservices} (MuFAno) project, we require a dataset of microservice monitoring data comprised of logs, metrics, and traces, to be studied jointly.
Given the challenge 3, we attempted to tackle the challenge 4 by leveraging the dataset of microservice projects \cite{amoroso2024dataset}
to identify a non-benchmark OSS microservice system to perform dynamic analysis and reconstruction, obtaining logs, metrics, and traces, thus providing a novel source of microservice monitoring data.
We performed dynamic analysis and testing of the seven microservices of the \emph{light-oauth2}\footnote{\url{https://github.com/networknt/light-oauth2}} project by \emph{networknt} \cite{bakhtin2025lo2}. However, authors of the project did not rely on an existing OpenTelemetry implementation
which required us to inject the Jaeger agent to gather the traces, resulting in almost null output.

Our attempts to find and describe the related work on microservice monitoring datasets also highlighted the fact that it is hard to find reliable datasets featuring all three modalities of monitoring data, i.e., logs, traces, and metrics \cite{bakhtin2025lo2}.
This presents yet another challenge:

\noindent\textbf{Challenge 5} \textit{Lack of substantial datasets of microservice monitoring data featuring traces, logs, and metrics.}

All in all, most encountered challenges are related to the lack of availability of reliable and representative microservice systems and related datasets.

\section{Conclusion}
In this research summary paper, I presented the ideas serving as the foundation of my research on applying temporal network methods for microservice architectural degradation. Moreover, I summarized the current status of its implementation and described several encountered challenges, most of which are due to the closed nature of big industrial players and the limited availability of OSS microservice benchmarks.

\bibliographystyle{splncs04}
\bibliography{bibliography}

\end{document}